\documentclass{jetpl}
\twocolumn
\title{Schwinger pair creation in multilayer  graphene}

\rtitle{Schwinger pair creation in multilayer  graphene }

\sodtitle{Schwinger pair creation in multilayer  graphene}

\author{M.A.Zubkov$^{*}$\/\thanks{e-mail: zubkov@itep.ru} }

\rauthor{M.A.Zubkov}

\sodauthor{M.A.Zubkov}

\address{
$^{*}$ ITEP, B.Cheremushkinskaya 25, Moscow, 117259, Russia
  }

\dates{April, 2012; preprint arXiv:1204.0138 }{*}


\abstract{The low energy effective field model for the multilayer graphene (at
ABC stacking) in external Electric field is considered. The Schwinger pair
creation rate and the vacuum persistence probability are calculated using the
semi - classical approach. }

\begin{document}

\maketitle

\section{Introduction}
\label{sectintr}

The Schwinger mechanism  of electron - positron pair creation in electric field
was first investigated in \cite{Schwinger1951}. The rate of pair production
depends on electric field and is so small that is not observable for the
experimentally allowed values of the electric field. In the condensed matter
systems the situation may be different. For example, in graphene
\cite{nature-438-197,Fialkovsky:2011wh} the large value of the effective
coupling constant opens the possibility for the pair creation process to be
observed \cite{allor2008,lew}. In graphene monolayer the pair creation rate may
be calculated using the approach of \cite{Schwinger1951}. However, a different
approaches were also used (see, for example,  \cite{allor2008,katsnelson2011}
and references therein). The approach described in \cite{katsnelson2011} was
also applied to the bilayer graphene \cite{Vildanov2009,volovik2012}.

In the present paper we calculate the pair production rate in multilayer
graphene. We consider the simplest case of ABC - stacking described by the two
- band pseudospin Hamiltonian with the chirality index $J$ equal to the number
of layers \cite{Multilayer}. In our calculations we rely on the method
developed in \cite{cohen2008,allor2008} and in \cite{gavrilov96}. A similar
approach was also applied to He-3  \cite{SV1992}. Within this approach, which
was originally applied to monolayer graphene, we develop the semi - classical
approximation (for the alternative ways to apply semi - classical technique to
the fermionic models see \cite{rajaraman,semiclass}). This approximation gives
results identical to the results obtained via the exact solution of the
Schrodinger equation for the case of monolayer. Our results obtained in the
multilayer graphene are checked with the more traditional semi - classical
approach described in \cite{katsnelson2011,Vildanov2009,volovik2012}.

The paper is organized as follows. In Section 2 we describe the one - particle
Schrodinger equation that appears in the given problem. In Section 3 we
introduce appropriate boundary conditions. In Section 4 the semiclassical
approximation for the one - particle Schrodinger equation is introduced. In
Section 5 we check the results obtained in Section 4 via the semiclassical
approach described in \cite{katsnelson2011,Vildanov2009,volovik2012}. In
Section 6 we compare our results with the exact ones for the case of graphene
monolayer. In Section 7 we calculate the vacuum persistence probability and the
pair production rate for the field - theoretic model of multilayer graphene. In
Section 8 we end with the conclusions.

\section{One - particle Schrodinger equation}

First, let us consider the one - particle problem. We deal with the two -
component spinors placed in the external Electric field directed along the $x$
- axis. We consider the external Electro-magnetic potential in the form: $A_x =
E t$. The one - particle Hamiltonian in a subsequent parametrization has the
form \cite{Multilayer,volovik2012}
\begin{equation}
H = v \left(\begin{array}{cc} 0 & \Bigl((\hat{p}_x + Et) - i \hat{p}_y\Bigr)^J \\
\Bigl((\hat{p}_x + Et) + i \hat{p}_y\Bigr)^J & 0\label{H1}
\end{array}\right)
\end{equation}

Here $v$ is a constant that is equal to Fermi velocity for the case of
monolayer. $J$ is the number of layers. Schrodinger equation has the usual form
\begin{equation}
i\partial_t \Psi = H \Psi
\end{equation}

Its solution is
\begin{equation}
\Psi(t) = P \,{\rm exp}\, \Bigl( -i  \int_{t_0}^t H(t)  dt\Bigr) \Psi(t_0) =
\hat{U}(t) \Psi(t_0)
\end{equation}
Here the path - ordered exponent is used. Operator $\hat{U}(t)$ is unitary by
construction. Later on we imply periodic boundary conditions in space -
coordinates. That's why $\Psi$ can be decomposed into the sum over the
quantized $2$ - momenta: $\Psi(t,x)) = \sum_{p_x,p_y} e^{i p_x x + i p_y y}
\psi_{p}(t)$. For $\psi(t)$ we have:
\begin{eqnarray}
\psi_p(t) &=&  P \,{\rm exp}\, \Bigl( -i v \int_{t_0}^t H[p,t] dt\Bigr) \psi_p(t_0), \\
H[p,t] & = & \left(\begin{array}{cc} 0 & \Bigl(({p}_x + Et) - i {p}_y\Bigr)^J \\
\Bigl(({p}_x + Et) + i {p}_y\Bigr)^J & 0
\end{array}\right)\nonumber
\end{eqnarray}

\section{Boundary conditions}

In our semiclassical consideration we imply that $J$ is odd. However,
analytical continuation will allow us to obtain final results for even values
of $J$ as well.  It is implied that at $t < t_0$ Electric field is absent and
we have
\begin{eqnarray}
\psi_p(t) &=& R[p,t]^+  e^{\,-i v |P|^J \sigma_3 t} R[p,t] \psi_p(t_0),\\
R[p,t] & = & \frac{1}{\sqrt{2}}\left(\begin{array}{cc} 1 & \Bigl(\frac{P^*}{|P|}\Bigr)^J \\
-\Bigl(\frac{P}{|P|}\Bigr)^J & 1\end{array}\right)\nonumber
\end{eqnarray}
where $P = p_x + E t_0 + i p_y$. Boundary conditions at $t_0$ must correspond
to the negative energy levels occupied:
\begin{eqnarray}
0&=&{\psi}_1 + \Bigl(\frac{P^*}{|P|}\Bigr)^J \psi_2, \nonumber\\
1 & = &\frac{1}{\sqrt{2}}[{\psi}_2 - \Bigl(\frac{P}{|P|}\Bigr)^J \psi_1]
\end{eqnarray}

It is supposed that at $t>t_0+T$ Electric field is switched off again.   Then
at $t = t_0 + T$ we have
\begin{eqnarray}
\eta_+&=&\frac{1}{\sqrt{2}}[{\psi}_1 + \Bigl(\frac{P^*}{|P|}\Bigr)^J \psi_2], \nonumber\\
\eta_-& = &\frac{1}{\sqrt{2}}[{\psi}_2 - \Bigl(\frac{P}{|P|}\Bigr)^J \psi_1]
\end{eqnarray}
the value $|\eta_+|^2$ is the probability that the electron - hole pair has
been created, while $|\eta_-|^2$ is the probability that the negative energy
level remains occupied.

For the semi - classical consideration it is useful to consider $t_0 = - T/2
\rightarrow -\infty$. The essence of semi - classical methodology is the
consideration of large frequencies that are in this case $p_x + Et$. That's why
we require at $t = t_0=-T/2$
\begin{eqnarray}
0&=&{\psi}_1 - \psi_2 \nonumber\\
1 & = &\frac{1}{\sqrt{2}}[{\psi}_2 + \psi_1].
\end{eqnarray}

At $t = +T/2$ we denote
\begin{eqnarray}
\eta_+&=&\frac{1}{\sqrt{2}}[{\psi}_1 + \psi_2] \nonumber\\
\eta_-& = &\frac{1}{\sqrt{2}}[{\psi}_2 - \psi_1],
\end{eqnarray}
where again the value $|\eta_+|^2$ is the probability that the electron - hole
pair has been created. This consideration implies $-ET/2+p_x<0$ and $ET/2+p_x
>0$.

For $ET/2  <  p_x $ ($p_x < - ET/2$) boundary conditions at $t = t_0=-T/2$ are:
\begin{eqnarray}
0&=&{\psi}_1 \pm \psi_2 \nonumber\\
1 & = &\frac{1}{\sqrt{2}}[{\psi}_2 \mp \psi_1].
\end{eqnarray}
Here the upper sign is for $ET/2  <  p_x $ while the lower one is for $-ET/2  >
p_x $. At $t = +T/2$ we expect
\begin{eqnarray}
\eta_+&=&\frac{1}{\sqrt{2}}[{\psi}_1 \pm \psi_2] \nonumber\\
\eta_-& = &\frac{1}{\sqrt{2}}[{\psi}_2 \mp \psi_1],
\end{eqnarray}
where again the value $|\eta_+|^2$ is the probability that the electron - hole
pair has been created.

\section{Semiclassical consideration}

Let us now introduce the notations:
\begin{eqnarray}
\tau &=& \Bigl(\frac{v}{E}\Bigr)^{\frac{1}{J+1}} (p_x+Et), \quad \Pi =
\Bigl(\frac{v}{E}\Bigr)^{\frac{1}{J+1}} p_y,\nonumber\\
 \quad \Theta & = & (\tau +
i \Pi)^J
\end{eqnarray}
Then
\begin{equation}
\psi_p(t) = P \,{\rm exp}\, \Bigl( -i \int \left(\begin{array}{cc} 0 & \Theta^* \\
\Theta & 0
\end{array}\right)d \tau \Bigr) \psi_p(t_0)
\end{equation}

The corresponding system of equations at $t > t_0$ is:
\begin{eqnarray}
i{\psi}^{\prime}_1 & = & \Theta^* \psi_2 \nonumber\\
i{\psi}^{\prime}_2 & = & \Theta \psi_1
\end{eqnarray}

For $\psi_{1,2}$ we have:
\begin{eqnarray}
&&\psi_2 = i{\psi}^{\prime}_1/u^J, \quad u = \tau - i \Pi \nonumber\\
&&[\frac{1}{u^J}\partial_u]^2 {\psi}_1 + (1+\frac{2i\Pi}{u})\psi_1
=0\label{psi1}
\end{eqnarray}

We introduce new variable $z = u^{J+1}/(J+1)$. The resulting equation is
\begin{eqnarray}
&&[\partial_z]^2 {\psi}_1 +
(1+\frac{2i\Pi}{(J+1)^{1/(J+1)}z^{1/(J+1)}})^J\psi_1 =0
\end{eqnarray}

We represent $\psi_1 = e^{is}$ and obtain the equation for $s$ that is
considered iteratively. In the first approximation we neglect the derivatives
of $s$ higher than the first derivative. In order to calculate the second
approximation we substitute the first approximation to the expression for
$s^{\prime\prime}$ etc. When only the first and the second terms are kept,  the
wave functions are given by
\begin{eqnarray}
\psi_1 &=&  c_1 \Bigl(\frac{\tau -i\Pi}{\tau+i\Pi}\Bigr)^{J/4} e^{-i
\int_{\tau_0}^{\tau} (\tau^2+\Pi^2)^{J/2}d\tau } \nonumber\\
&&+ c_2 \Bigl(\frac{\tau -i\Pi}{\tau+i\Pi}\Bigr)^{J/4} e^{i
\int_{\tau_0}^{\tau} (\tau^2+\Pi^2)^{J/2}d\tau }
\nonumber\\
\psi_2 & = &  c_1 \Bigl(\frac{\tau +i\Pi}{\tau-i\Pi}\Bigr)^{J/4} e^{-i
\int_{\tau_0}^{\tau} (\tau^2+\Pi^2)^{J/2}d\tau } \nonumber\\
&&- c_2 \Bigl(\frac{\tau +i\Pi}{\tau-i\Pi}\Bigr)^{J/4} e^{i
\int_{\tau_0}^{\tau} (\tau^2+\Pi^2)^{J/2}d\tau }
\end{eqnarray}

The considered approximation is valid if the second approximation is smaller
than the first one. This leads to the condition
\begin{equation}
|\frac{J \Pi}{(\tau^2 +
\Pi^2)^{J/2+1}}|=\frac{p_y\Bigl(\frac{E}{v}\Bigr)^{\frac{J}{2(J+1)}}}{\Bigl((p_x+Et)^2
+ p_y^2\Bigr)^{\frac{J+2}{2}}}<<1
\end{equation}

 Next, we use the fact that the
given semi - classical approximation gives the solution of Eq. (\ref{psi1}) not
only for large real values of $\tau$ but for the complex values of $\tau$ with
large $|\tau|$. That's why analytical continuation of the solution at $\tau
\rightarrow -\infty$ gives the solution at $\tau \rightarrow +\infty$. (The
continuation is performed along the line placed at $|\tau|\rightarrow \infty$.)

At $ET/2  >  p_x >  -ET/2$ boundary conditions give $c_2 = 0$. The probability
that the pair is created is $|\eta_+|^2$, where
\begin{eqnarray}
\eta_+ & = & e^{-i \int_{\tau_0}^{\tau} (\tau^2+\Pi^2)^{J/2}d\tau} =
e^{\frac{|\Pi|^{J+1}}{2} \int (1-z)^{J/2}z^{-1/2}dz} \nonumber\\ &=&
e^{-|\Pi|^{J+1} B(J/2+1,1/2)}, \quad z = -(\tau/\Pi)^2
\end{eqnarray}
Here we consider the contour placed at infinity with the orientation such that
$|\eta_+|$ remains less than unity.

We have
\begin{eqnarray}
|\eta_+|^2 & = &  e^{-\alpha \Bigl(|p_y|/E^{1/(J+1)}\Bigr)^{J+1}}, \nonumber\\
\alpha & = & 2 \pi  v \frac{ J!!}{(J+1)!!} = 2 v
B\Bigl(\frac{J}{2}+1,\frac{1}{2}\Bigr) \nonumber\\
&=& 2 v J B \Bigl(\frac{3}{2}, \frac{J}{2}\Bigr)\label{alpha}
\end{eqnarray}

Written in this form our result can be continued analytically to even values of
$J$. At the same time known results for $J = 1,2$ are reproduced
\cite{allor2008,volovik2012}.

At $ET/2  <  p_x $ ($p_x < - ET/2$) boundary conditions give $c_1 = 0$
($c_2=0$). In both cases semicassical approximation gives $|\eta_-|=1$ that
means that the electron - hole pair is not created.


Below we check the obtained above value of the probability that the electron -
hole pair is created with the given values of momenta $(p_x,p_y)$. We do this
in two ways: via the application of the semi - classical approximation in its
more classical form and via the consideration of the exact solution of the
Schrodinger equation (at $J = 1$).

\section{More classical semi-classics}

The problem of pair creation can be considered in the gauge $A_0 = -Ex$. Then
we have stationary Schrodinger equation $H \Psi = \epsilon \Psi$ with
\begin{equation}
H =  \left(\begin{array}{cc} E x & v\Bigl(\hat{p}_x - i \hat{p}_y\Bigr)^J \\
v\Bigl(\hat{p}_x + i \hat{p}_y\Bigr)^J & E x
\end{array}\right)
\end{equation}
We proceed with the rescaling $z = \Bigl(\frac{E}{v} \Bigr)^{1/(J+1)} x$, and
$\omega = \Bigl(\frac{1}{v E^J} \Bigr)^{1/(J+1)}\epsilon $. Then
\begin{eqnarray}
&&(z-\omega) \psi_1 + (-i\partial_z - i \Pi)^J \psi_2=0 \nonumber\\
&&(z-\omega) \psi_2 + (-i\partial_z + i \Pi)^J \psi_1 =0
\end{eqnarray}

The first order semi - classical approximation for $\psi_{1,2}$ gives
\begin{eqnarray}
&&\psi =   e^{\pm i \int (-\Pi^2+(z-\omega)^{2/J})^{1/2}d z }
\end{eqnarray}
Integration over the classically forbidden region $-\Pi^2+(z-\omega)^{2/J} < 0$
gives the pair production probability:
\begin{eqnarray}
|\eta|^2 & = & e^{-\alpha \Bigl(p_y/E^{1/(J+1)}\Bigr)^{J+1}}, \nonumber\\ &&
\alpha = 2  v J B\Bigl(\frac{3}{2}, \frac{J}{2}\Bigr)
\end{eqnarray}
This expression coincides with the one derived above.

\section{Exact solution at $J = 1$}

Let us introduce notations $\psi_+ = \psi_1 - \psi_2$ and $\psi_- = \psi_1 -
\psi_2$. Then
\begin{eqnarray}
&& \psi_-  =  \frac{1}{\Pi} {\psi}^{\prime}_+ + i \frac{\tau}{\Pi} \psi_+ \nonumber\\
&& {\psi}^{\prime\prime}_+ + (i + \Pi^2 + \tau^2) \psi_+ = 0
\end{eqnarray}

We change variables $\tau = \frac{1}{\sqrt{2}}e^{i\pi/4} z$. Then
\begin{eqnarray}
&& \psi_-  =  \frac{\sqrt{2}e^{-i\pi/4}}{\Pi}[\partial_z-z/2]{\psi}_+ \nonumber\\
&& {\psi}^{\prime\prime}_+ + (1/2 + [\frac{i\Pi^2}{2}-1] - z^2/4) \psi_+ = 0
\end{eqnarray}
The solution is
\begin{eqnarray}
 \psi_+  &=& b_1 D_{-i\Pi^2/2}(\sqrt{2}e^{-3i\pi/4}\tau) + b_2 D_{i\Pi^2/2-1}(\sqrt{2}e^{3\pi i/4}\tau)  \nonumber\\
 \psi_-  &=& \frac{\sqrt{2}e^{-i\pi/4}}{\Pi}\{-b_1 \Pi^2/2
D_{-i\Pi^2/2-1}(\sqrt{2}e^{-3i\pi/4}\tau)\nonumber\\
&& + b_2 D_{i\Pi^2/2}(\sqrt{2}e^{3i\pi/4}\tau)\}
\end{eqnarray}

The consideration of usual boundary conditions leads to rather complicated
algebra. So we come to the semiclassical boundary conditions ($\tau_0
\rightarrow   + \infty$):
\begin{eqnarray}
 \sqrt{2} & = & b_1 D_{-i\Pi^2/2}(\sqrt{2}e^{i\pi/4}\tau_0) + b_2 D_{i\Pi^2/2-1}(\sqrt{2}e^{-i\pi/4}\tau_0)  \nonumber\\
 0 & = &   -b_1 \Pi^2/2 D_{-i\Pi^2/2-1}(\sqrt{2}e^{i\pi/4}\tau_0)\nonumber\\
&& + b_2 D_{i\Pi^2/2}(\sqrt{2}e^{-i\pi/4}\tau_0)
\end{eqnarray}

Using asymptotic expansion for Weber function we come to $b_2=0$ and
\begin{equation}
|\eta_+|^2 = e^{-\pi\Pi^2}=e^{-\pi\Bigl(\frac{v_F}{E}\Bigr)p_y^2}
\end{equation}
is the probability that the electron - hole pair is created.

\section{Field - theoretical consideration}

The fact that the particles do not interact with each other allows to reduce
the field - theoretical problem to the quantum mechanical one. Namely, we
arrive at the following pattern. Modes for different values of momenta
propagate independently. At $t \le t_0$ all states with negative values of
energy are occupied while all states with positive values of energy are vacant.
Their evolution in time is governed by the one - particle Schrodinger equation.
At $t = t_0 + T$ the wave function already has the nonzero component
corresponding to positive energy. Its squared absolute value is the probability
that the electron - hole pair is created.

Let us calculate the probability that vacuum remains vacuum $P_v$ (vacuum
persistence probability). According to the above presented calculation this
probability is
\begin{equation}
P_v = \Pi_{p_x,p_y} \Bigl(1-e^{-\alpha
\Bigl(|p_y|/E^{1/(J+1)}\Bigr)^{J+1}}\Bigr)^{g_s g_v} = e^{-2 {\rm Im} S}
\end{equation}
 Here $S$ is the effective action, the factors $g_s=2$ and $g_v = 2$ are spin and valley degeneracies.  The product is
over the momenta that satisfy
\begin{equation}
ET/2  >  p_x > -ET/2
\end{equation}

 We have ($L$ is the linear size of the graphene sheet):
\begin{eqnarray}
\omega & = & \frac{2\, {\rm Im} S}{TL^2}  = - g_s g_v \frac{E}{2\pi L}
\sum_{p_y = \frac{2\pi}{L} K} {\rm log} (1-e^{-\alpha
\Bigl(|p_y|/E^{1/2}\Bigr)^{J+1}})\nonumber\\ &\approx& - g_s g_v\frac{E}{2\pi}
\int \frac{dp_y}{2\pi} {\rm log} (1-e^{-\alpha
\Bigl(|p_y|/E^{1/(J+1)}\Bigr)^{J+1}})\nonumber\\
&=&  g_s g_v \frac{E}{2\pi} \sum_n \frac{1}{n} \int \frac{dp_y}{2\pi}
e^{-\alpha n \Bigl(|p_y|/E^{1/(J+1)}\Bigr)^{J+1}}\nonumber\\& =&   g_s
g_v\frac{ E^{\frac{J+2}{J+1}}}{2(J+1)\pi^2(\alpha)^{1/{(J+1)}}}
\zeta\Bigl(\frac{J+2}{J+1}\Bigr)\Gamma\Bigl(\frac{1}{J+1}\Bigr)
\end{eqnarray}

Here $\alpha$ is given by Eq. (\ref{alpha}).  According to \cite{cohen2008} a
different quantity is considered as the pair production rate:
\begin{eqnarray}
\Gamma & = & \langle |\eta_+|^2 \rangle/(L^2T)  = g_s g_v \frac{E}{2\pi L}
\sum_{p_y = \frac{2\pi}{L} K} e^{-\alpha \Bigl(|p_y|/E^{1/2}\Bigr)^{J+1}}
\nonumber \\ &=&  g_s g_v \frac{
E^{\frac{J+2}{J+1}}}{2(J+1)\pi^2(\alpha)^{1/{(J+1)}}}
\Gamma\Bigl(\frac{1}{J+1}\Bigr)
\end{eqnarray}

The form of the functional  dependence of $\Gamma$ on $E$ coincides with that
of mentioned in \cite{volovik2012}.

\section{Conclusions}
\label{sectconcl}

In the present paper we calculate the pair production rate and the vacuum
persistence probability for the multilayer graphene. We develop the
semiclassical technique within the approach used earlier in monolayer graphene.
Our method reproduces known results for monolayer and bilayer graphene.
Following \cite{cohen2008} we consider the single pair creation rate $\Gamma$
and $\omega = - \frac{{\rm log} P_v}{TL^2}$ (where $P_v$ is the vacuum
persistence probability) as different quantities. The possibility to consider
$\omega$ as a production rate of multiple states remains open and requires an
additional investigation.

The author kindly acknowledges discussions with G.E.Volovik. This work was
partly supported by RFBR grant 11-02-01227, by Grant for Leading Scientific
Schools 6260.2010.2, by the Federal Special-Purpose Programme 'Cadres' of the
Russian Ministry of Science and Education, by Federal Special-Purpose Programme
07.514.12.4028.

\end{document}